\documentclass[preprint,aps]{revtex4}
\usepackage{amssymb,amsmath}
\usepackage{graphicx,float}

\tighten
\begin{document}

\title{ \Large \bf \flushleft 
 Information Super-Diffusion on Structured Networks}
\author{\large  \flushleft 
Bosiljka Tadi\'c$^{1}$ and 
Stefan Thurner$^{2}$
} 

\affiliation{ \flushleft
$^1$Department  for Theoretical Physics, Jo\v{z}ef Stefan Institute; 
P.O. Box 3000; SI-1001 Ljubljana; Slovenia, 
$^2$Institut f\"ur Mathematik NuHAG and HNO AKH, Universit\"at Wien; 
W\"ahringer G\"urtel 18-20; A-1090 Vienna;  Austria  \\ \hspace{1cm}
} 

%\date{}
\begin{abstract}
\noindent
We study diffusion of information packets 
on several classes of structured networks. 
Packets diffuse from a randomly chosen node 
to a specified destination in the network. 
As local transport rules we consider random 
diffusion and an improved local search method.
Numerical simulations are performed in the regime 
of stationary workloads away from the jamming 
transition.  We find that graph topology determines 
the properties of diffusion in a universal way, 
which is reflected by power-laws in the transit-time 
and velocity distributions of packets. 
With the use of multifractal scaling analysis and 
arguments of non-extensive statistics we find that 
these power-laws are compatible with super-diffusive 
traffic for random diffusion and for improved local search.  
We are able to quantify the role of network 
topology  on overall transport efficiency. Further, we  
demonstrate the implications of improved transport rules 
and  discuss the importance of matching (global) topology with 
(local) transport rules for the optimal function of networks.
The presented model should be applicable to a wide range of 
phenomena ranging from Internet traffic to protein transport 
along the cytoskeleton in biological cells.  
\vspace{0.5cm}

\end{abstract}

\maketitle

\section{Introduction}
Properties of different types of networks 
have begun to attract the interest of statistical physicists. 
The spectrum of network related problems include, among many others,  
ordinary traffic in a city \cite{cartraffic}, distribution of nutrients 
in the vascular system \cite{west}, distribution of goods and wealth 
in economies, queuing problems on information networks
\cite{traffic,queuing},
biochemical- and gene expression  pathways \cite{genetic_nets}. 
It has been recognized that most natural and man-made networks evolve 
in time and that their structure emerged as result of microscopic
evolution rules \cite{SD}. In many networks a scale-free linking structure  
was found \cite{SD,AB} which emerges  
from complex processes of self-organization with specific constraints. 
Maybe the most prominent example of such a scale-free network is  
the world-wide Web \cite{Web, BT} and the Internet
\cite{Internet_structure}.
Transport processes on networks, such as traffic of  information packets on 
the Internet or diffusion of signaling molecules in a biological  cell, are 
physical processes which are closely related to the network geometry. 
It is likely that biological networks have  been 
optimized in structure by selections in evolution. 
The vast research on information traffic on technological networks is 
motivated to optimize network structure for 
practical reasons such as, e.g., optimum connection strategies 
and information flow at minimal costs and risk. 
Technological networks like the 
Internet or the world-wide Web  are relatively easily accessible 
for empirical research in contrast to e.g. biological networks 
such as metabolic pathways.
A further important motivation for studying networks is to gain theoretical
insight in the role of an underlying geometry on the function 
of a network. 

Recent empirical studies on Internet traffic suggest a complex interplay of 
queuing behavior  \cite{queuing}, temporal correlation in packet streams 
\cite{erramilli,t1}, and broad distributions of travel times of packets 
\cite{t1}. 
Moreover, evidence is gathered that both traffic parameters (traffic intensity,
buffer sizes, link capacities)  
and network connectivity play a role in the overall network's performance. 
Often two general regimes of traffic are recognized in networks: free flow 
and jammed traffic. A jamming transition is expected when
traffic density exceeds a critical value. 

In numerical work recently a model of packet transport was 
introduced \cite{TR} which studies 
simultaneous random walks to specified destinations on scale-free networks. 
The results suggest that fat-tailed transit-time distributions  and temporal
correlations may occur for a wide range of values of packet densities as a 
consequence of network structure \cite{TR}. 
Other theoretical models were  able to capture  essential properties of the 
jamming transition on simple network geometries, like hierarchical trees 
\cite{Albert_ht} and square lattice models \cite{Albert_sq,Sole1,Sole2}.

In this work we implement a general numerical model 
to study packet traffic on several
classes of networks with diverse structural characteristics. These 
include a highly organized cyclic scale-free graph and two types
of tree-like graphs. 
By working in a stationary traffic regime below the jamming transition, 
our aim is to relate a given network topology 
to characteristic  diffusion parameters such as 
probability distributions of travel times and packet velocities. 
Further we want to quantitatively characterize the diffusion on given network 
topologies within a general framework of non-linear diffusion phenomena.

Quite generally, transport on networks can be seen as a  
diffusion on a geometry which is not plane- or space filling. 
Diffusion on such topologies has been a matter of interest over the 
past decades. Many problems like the diffusion 
of water particles in a sponge, or crude oil in porous media 
are phenomenologically well described by the anomalous diffusion 
equation.   
Anomalous diffusion is naturally linked to 
power laws in the corresponding spatio-temporal probability 
distributions.  
It has been shown that the probability density which 
maximizes entropy of a version of a non-extensive 
entropy definition \cite{tsallis88} is an exact solution 
of the anomalous diffusion 
equation \cite{tsallis96}.  
It is part of this work to notice the connections 
between transport on networks and anomalous diffusion and 
to use the latter to quantify topology-dependence of network performance.

Queuing processes on networks are currently subject of intensive study. 
Mathematically, being composed  of many random processes, network queues 
are a natural  laboratory to study limit stochastic processes. The  
existence of stable laws in these processes is expected \cite{book_spl}. 
However, the connection  between the character of the limit 
process and network structure remains an open theoretical problem.
Having numerically determined large  series of transit-times of packets 
 we also attempt to recognize their convergence to a stable law  and  
explore the consequences of it on specific network topologies.

In Section \ref{secii} we specify the model of network diffusion 
and describe details of network structures and the implementation 
of diffusion rules. We determine conditions of the 
stationary (un-jammed) regime for all network topologies.
In Section \ref{seciii} the results for transit-times distributions
are presented in the stationary
regime. We show results for the limit of infinitely low packet density on 
networks and discuss the statistical regularity of transit-timeseries. 
Section \ref{seciv} motivates  the use of anomalous diffusion methods 
in network transport and contains the results of the velocity
distributions on different networks. In Section \ref{secv} we study temporal
correlations of workload timeseries and in 
Section \ref{secvi}  we summarize and discuss the results.

\section{Model of Non-linear Diffusion on Networks \label{secii}}
A fundamental problem in non-linear diffusion is
to understand how the topological structure or the sparseness of 
space affects the dynamics that it supports.
Transport on a complex network with local navigation rules is
a  prototype for  diffusion on a sparse structure. 
Further reasons for non-linear terms in diffusion equations can 
be related  to the diffusion process itself, 
i.e.,  to the specific way of local navigation between individual nodes and 
queuing properties.
In this work we introduce a numerical model in which we can control both, 
graph topology and microscopic diffusion details (transport rules). 
We consider graphs with link structure that emerges from  
controlled microscopic linking rules, from which the graphs are grown.  
We grow three different types of networks
(described in detail below) consisting of an equal   
number of nodes ($N=1000$) and links ($E=1000$) but having 
different linking patterns. 
After specifying the diffusion rules we monitor the {\it simultaneous} 
transport of packets on these graphs,  
and analyze local and statistical properties of the resulting traffic.
A packet (particle) carries information from the node where it is created
to an in-advance specified node (address) on the network.
 These can be, for instance,  a piece of information 
transported on the Internet or a protein in a biological cell.

\subsection{Types of networks}
To determine how the structure of links in the network influences  
diffusion, we study three topologies with emergent link structures.
Their main  structural characteristics are  the following: 
(i) Scale-free directed graph with closed cycles (cyclic Web graph, WG); 
(ii) Scale-free graph with no closed cycles (scale-free tree, SF); 
and (iii) Absence of a scale-free structure and absence of
closed cycles (randomly grown tree, RG).

For all cases we stop growth of the graph when  
$N=E=1000$  nodes and links have been added.
All the graphs that we consider have the same number of nodes and 
links, they differ in the adjacency matrix. 
After growing the network, the corresponding adjacency matrix is not 
changed during the diffusion processes. 
   
\subsubsection{The Web graph (WG) as a scale-free cyclic graph}
\begin{figure}[htb]
\begin{center}
\begin{tabular}{cc} 
\includegraphics[width=7.5cm]{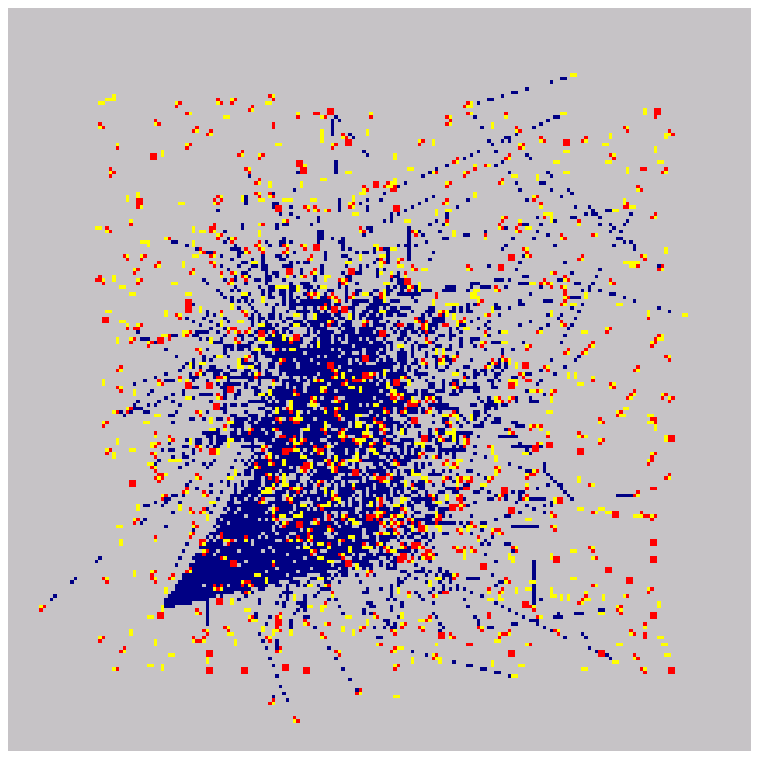}&
\includegraphics[width=7.5cm]{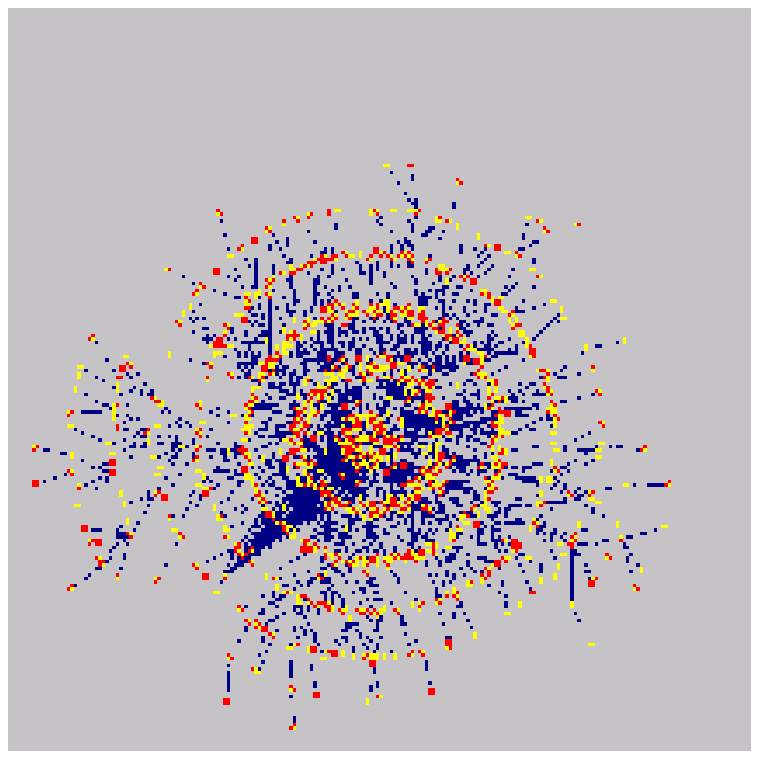} \\
{\Large $(a)$} &{\Large $(b)$}  \\
\end{tabular}
\end{center}
\caption{
(a) Web graph (WG) and (b) scale-free tree graph (SF) composed  of 
the same number of nodes and links, $N=E=1000$. The graphs are 
grown from the microscopic rules described in the 
text and in Ref. \cite{BT}. 
The first hub node 
is pulled towards the lower left corner for better vision. In the WG 
an additional hub is visible in the central part of the figure. 
Typical layers of nodes linked to  preceding layers are seen as 
circular arrangements in the SF,
which are absent in the WG. Disconnected nodes occurring in the 
WG have a finite probability to increase both in- and out-link 
connectivity if the graph continues to grow, while in the SF 
the out-link connectivity of the nodes remains fixed. }
\label{figgraph}
\end{figure}

We first consider the scale-free cyclic graph (WG) 
shown in Fig. \ref{figgraph}a.  
Such a graph is grown by adjusting a single free parameter 
governing microscopic linking rules proposed in \cite{BT}. 
These rules have been 
suggested as an evolution model of the world-wide
Web. The minimal set of rules consists of: 
{\it growth} by adding of a new node at each time step and  
{\it attachment} (with probability 
$\tilde{\alpha}$) of the new node to a pre-existing node,  or 
{\it rewiring} (with probability
$1- \tilde{\alpha }$) of a pre-existing node.  
Each link is oriented originating from a node, $n$,
as an out-going link,  and pointing to another node, $k$, where it is seen 
as in-coming link. In the model  \cite{BT,BT_ccp02} both 
the target node $k$ and and the origin $n$ of a link
are selected  by shifted preferential probabilities given by 
\begin{equation}
p_{in}=  (M\alpha + q_{in}(k,t))/(1+\alpha )Mt \   ;   \
p_{out}= (M\alpha + q_{out}(n,t))/(1+\alpha )Mt \ ,
\label{Pt}
\end{equation}
which depend on the  number of respective in-coming links $q_{in}(k,t)$,
and  out-going links $q_{out}(n,t)$ at current time $t$.
The shift parameter $\alpha$ is the initial probability, where $q_{in}(k,0)=
q_{out}(n,0) =0$. $M$ is the average number of links {\it added} per time step.
As shown in \cite{BT,BT_ccp02} the biased attachment and rewiring in the 
growth rules lead to a  scale-free organization of both in- 
and in out-links with the local connectivity at node $s$ at time $t$ 
building up in time with a power-law 
\begin{equation}
q_{in} (s,t) = A_1\left[\left(t/s\right)^{\gamma _{in}}  
-B_1 \right] \ ; q_{out} (s,t) = A_2\left[\left(t/s\right)^{\gamma _{out}}  
-B_2 \right]
\label{local_connectivity}
\end{equation}
where  the scaling exponents are
$\gamma _{in} = 1/(1+\alpha )$,  and 
$\gamma _{out} = (1-{\tilde{\alpha }})/(1+\alpha )$.  
The power-law growth of the local connectivity in Eq.\ (\ref{local_connectivity})
ensures an emergent scale-free structure of the graph 
after long evolution time. The global graph structure at 
$t\to \infty$ is thus characterized by the power-law 
distributions of links 
\cite{BT,BT_ccp02,SD},
$P(q_{in})\sim q_{in}^{-\tau_{in}}$ 
and  $P(q_{out})\sim q_{out}^{-\tau_{out}}$ with  
$\tau_{in} =1 + \gamma_{in}^{-1}$ and 
$\tau_{out} =1 + \gamma_{out}^{-1}$.
Here we use the original one-parameter model \cite{BT} with 
$\alpha =\tilde{\alpha } =1/4$ and $M=1$.
Apart from the scale-free organization of in- and out-links, 
the emergent graph 
structure is characterized by the occurrence of closed 
cycles of directed links and a single giant connected component.  
Two types of hubs (nodes with a large in-link and nodes with a large
out-link connectivity) appear together with a substantial 
degree of correlations 
between local in- and out-connectivity \cite{BT_ccp03}. 
Recent analysis of empirical data on the structure of the Internet 
\cite{Internet_structure}
reveal similar topological properties on the level of routers.
This type of graph topology has also importance in biological scale-free
networks (metabolic networks) 
\cite{cell,metabolic_nets}, where the presence of closed cycles 
enables feedback regulation. As a further example we mention  
gene co-regulation networks \cite{genetic_nets}. After a gene is 
expressed its product molecules (mRNA or proteins) are in part 
addressed to 
specific genes in the network where they bind and regulate the 
expression of the other genes. 
Another  biological network which shares some characteristics 
of the WG is the cell crawling network, where cells 
move along dynamic paths traced by previous cells \cite{mombach,Stefan}. 

\subsubsection{The scale-free tree graph (SF)}
To grow the scale-free tree graph (SF) 
we use the shifted preference rule with
$p_{in}$ given in Eq.\ (\ref{Pt}), but we suppress rewiring of links 
by fixing the parameter $\tilde{\alpha } =1$, and set 
the parameter $M=1$.
Therefore each added node connects immediately to a  
pre-existing one, which is  selected with a probability $p_{in}$ in 
Eq.\ (\ref{Pt}), and the link remains fixed in the subsequent evolution
 of the graph.
For a characteristic look of such a graph, see Fig. \ref{figgraph}b.  
For better comparison, for the SF graph we chose
the shift parameter $\alpha =1/4$, so that the   
in-link connectivity is described   
by the same power $\gamma _{in}$ as the WG in 
Eq.\ (\ref{local_connectivity}). However, the out-links 
remain evenly distributed, i.e., one link per node, hence making the overall
graph structure less organized compared to the WG. In particular, only a hub 
with large in-link connectivity can occur.
When the number of links added per step  $M$ exceeds one the tree structure 
is lost and the graph can  have multiple paths from one node to another. 
These may be considered as cycles if the 
directedness of the links is disregarded, in contrast to the WG where closed 
cycles along strictly directed links occur even for $M=1$. Another 
structural difference of the SF and the WG is the absence of 
correlations between local in- and out-connectivities in the SF. 
The SF (with general $M$) is most often called scale-free network 
in the literature \cite{AB,SD,TR}.

\subsubsection{The randomly grown tree  graph (RG)}
The randomly grown tree graph is grown in the same way as the SF tree except 
that the target node for each link is selected randomly among all the 
pre-existing nodes. At time  $t$ the probability 
to link to a selected  node
is $p_{in}(t) = t^{-1}$. This linking rule leads to a local 
connectivity increase which is logarithmic 
in time rather than a scale-free law. 
The out-link connectivity remains fixed with 
one link per node. In the present context the RG is the least organized 
or structured graph.

\subsection{Diffusion rules on the network}
Given the graph structure in form of the adjacency matrix, 
diffusion of packets is defined by specifying the following 
steps (see also  \cite{TR,BT_ccp03}):
\begin{itemize}
\item{} {\it Creation and assignment}: 
A packet is created with rate $R$ (posting rate)
at a randomly selected node and it is 
assigned a destination address on the graph. 

\item{} {\it  Diffusion rules}: Before reaching its destination a 
packet passes from a node to one of its neighbors,
 being directed by local search algorithms 
(diffusion rules). 
For the selection of a neighbor we implement two algorithms: 

(1) {\it Random diffusion} (RD): The packet is passed from a given node 
to one of its neighbors 
with {\it equal} probability; 

(2) {\it Cyclic Search} (CS): The node performs a 
next-to-nearest-neighbor (nnn) search in its 
neighborhood, looking if 
the destination address of the packet is within nnn distance.  
If it gets a signal that the address was 
found within a nnn surrounding, the node passes the packet to that neighbor.
If the address is not within nnn it uses the random diffusion rule (1). 

\item{} {\it Queuing strategy}: When more than one packet 
is sitting on a 
node a queue is formed and a {\it queuing discipline} needs to be 
specified. Here we mainly focus on the first-in-first-out (FIFO) queue, 
but also study the last-in-first-out (LIFO) strategy. 
A packet can only be processed if the current queue size at  a 
selected neighbor node is bellow some maximal allowed buffer size 
$H$, otherwise
it waits for the next occasion to be processed. For simplicity, we treat 
diffusion along out-links and against in-links with equal probability. 
The model allows for various other possibilities.

\item{} {\it Removal}: When a packet  reaches its destination it 
will be removed from the network. 
\end{itemize}

To implement these steps one first defines a 
time step by one {\it parallel} update of the entire network. 
At each time step we create  packets with rate $R$ and assign them their 
destination addresses. In order to avoid 
sending packets to a disconnected part of the graph we specify the list  of 
nodes belonging to the giant component of the WG. The  pairs of nodes 
(origin and destination) is selected randomly 
from that list. The same list is  used for all three  graph types.
At each time step we also keep track of 
the packets' current positions, and their positions 
within a queue at given node. 
For temporal traffic statistics we label a certain  
number of packets so that we  monitor  their
creation time, the time that a packet spent at a current node, and 
the total elapsed time before being delivered to its destination.
In our setup we label 2000 packets.  
The length of a particular 
simulation for a given network and search algorithm depends on the 
posting rate $R$.
We keep updating the system until the pre-fixed number  
of labeled  packets have been created and arrived at their destinations.  
Results are averaged over many network realizations.  We typically use 
100 sample networks for each simulation. Details of the numerical 
implementation of the code can be found in Ref. \cite{BT_ccp03}.

\subsection{Stationary Flow and Slow Driving}

\begin{figure}[htb]
\begin{center}
\begin{tabular}{c} 
\includegraphics[width=12.5cm]{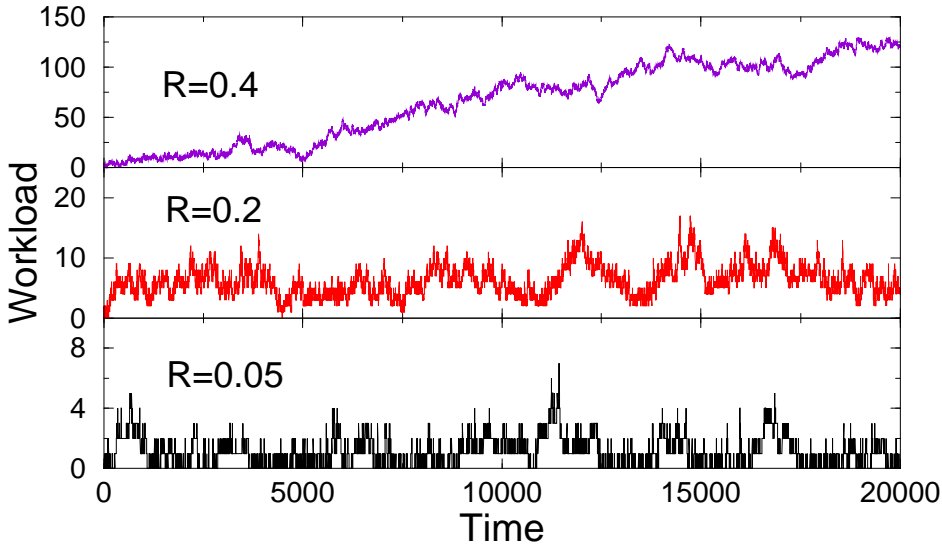} 
\end{tabular}
\end{center}
\caption{Number of packets in the WG network as a function of time 
(workload) for the advanced search algorithm (CS) and different 
posting rates $R=0.05$, $R=R^{\star}=0.2$, and $R=0.4$. 
}
\label{figrate}
\end{figure}

To be able to entirely focus on the network structure 
and not to be troubled with queuing effects at 
individual nodes, we assume a large buffer $H=1000$ at each node
and apply a posting rate $R$ such that the number of 
packets present in the system fluctuates around a
finite value considerably below the maximal queue size $H$.   
An example of stationary flow is shown in lower two panels in Fig. 
\ref{figrate}.
Stationarity is established by the balance of the 
network's input rate $R$ and its output rate $\rho $, 
i.e.,  the number of delivered packets
per time step. The output rate $\rho$ is 
a parameter which depends on the network topology and the 
diffusion rules. 
The number $n(t)$ of packets in the system at 
current time $t$ and is called {\it workload}.  
When the posting rate exceeds a certain limit, 
the workload $n(t)$ shows a persistent increase, 
which is related to the occurrence of 
jamming (full buffers) in the network. In fact for several networks 
there exists a phase transition from a free-flow phase to a 
jammed phase \cite{Sole1}. Here we do not discuss the onset of jamming, 
our aim is to study properties of traffic on different networks 
in  stationary flow below the jamming transition. 
Values of the posting rate $R$ for which a stationary flow 
is sustainable appear to be crucially dependent both on 
network topology and search algorithms. 
To be able to compare the efficiency of a 
given network, we select to work with a posting rate $R^\star$,
for which the workload fluctuates within same
limits when different search algorithms are used. 
As a rule, when the rate is increased above $R^\star$, 
jamming will first occur in the case of less efficient search. The values 
for $R^\star$ for two search algorithms  in the three types of graphs  
are summarized in Table \ref{table_1}. 
\begin{table}[H]
\begin{center}
\begin{tabular}{|c|c c c|} \hline    
                  & $R^\star$ (CS) &~~~ $R^\star$ (RD)  & ~~~ Ratio CS:RD \\ 
\hline
          WG      & $ 2\times 10^{-1}$&~~~ $ 5\times10^{-3} $&  $ 40 $\\ 
%\hline
          SF      &  $5\times10^{-3} $ &~~~  $1\times10^{-3} $ &  $ 5 $\\ 
%\hline
          RG      &  $1\times10^{-2 }$ & ~~~ $5\times10^{-3} $ &  $  2 $\\ 
\hline
\end{tabular}
\end{center}
\caption{Values of posting rates  $R^\star$  
for the search algorithms CS and RD and different graphs.  
The ratio CS:RD shows the efficiency increase of the 
CS relative to RD for given  graph topology.}
\label{table_1}
\end{table} 
The improvement of efficiency of the 
advanced local nnn-search relative to random diffusion is shown
for the different topologies in the column ''Ratio CS:RD''.
In the stationary regime within a given time 
interval the cyclic WG manages to process 40 times more packets by using 
CS than when using RD. The same search 
algorithm is doing five times better on the SF tree,  and 
only a factor two on the RG.

As a limiting case where queuing is completely absent we define the 
{\it zero posting rate} limit, $R\to 0$, where a new packet is created only 
{\it after} the previous one is delivered at its destination. 
This means that there is only one packet in the system at any time. 
Although this limit is not accessible in real network traffic, 
it is mathematically 
interesting because the absence of queuing allows to 
single out topology  effects on non-linear diffusion. 

\section{Transit times on different graph topologies \label{seciii} } 

Transit time $T$ is the time a packet spends on the network
from posting until arrival at its destination. 
It is primarily related to the number of jumps the packet performs 
on the graph which is crucially dependent on the 
topology and the search algorithm. For a finite posting rate $R>0$  
the effects of queuing also contribute to the transit-time.  For a given 
packet  the transit-time is determined as 
 \begin{equation}
T_k = \sum _{i=1}^k t_w(i) \quad,
\label{Tk} 
\end{equation}
where index $i$ denotes nodes along the actual path of the packet, 
and $t_w(i)$ is the packet's waiting-time at node $i$ along the path. 
$k$ is the total length of the path taken by the packet. 
We determine the probability distribution 
of transit-times, $P(T)$, in the stationary flow regime. 
$P(T)$ incorporates two effects which are crucial for the 
diffusion on networks: the graph topology and waiting-times.

\subsection{The $R\to 0$ limit}
\begin{figure}[htb]
\begin{center}
\begin{tabular}{c} 
\includegraphics[width=13cm]{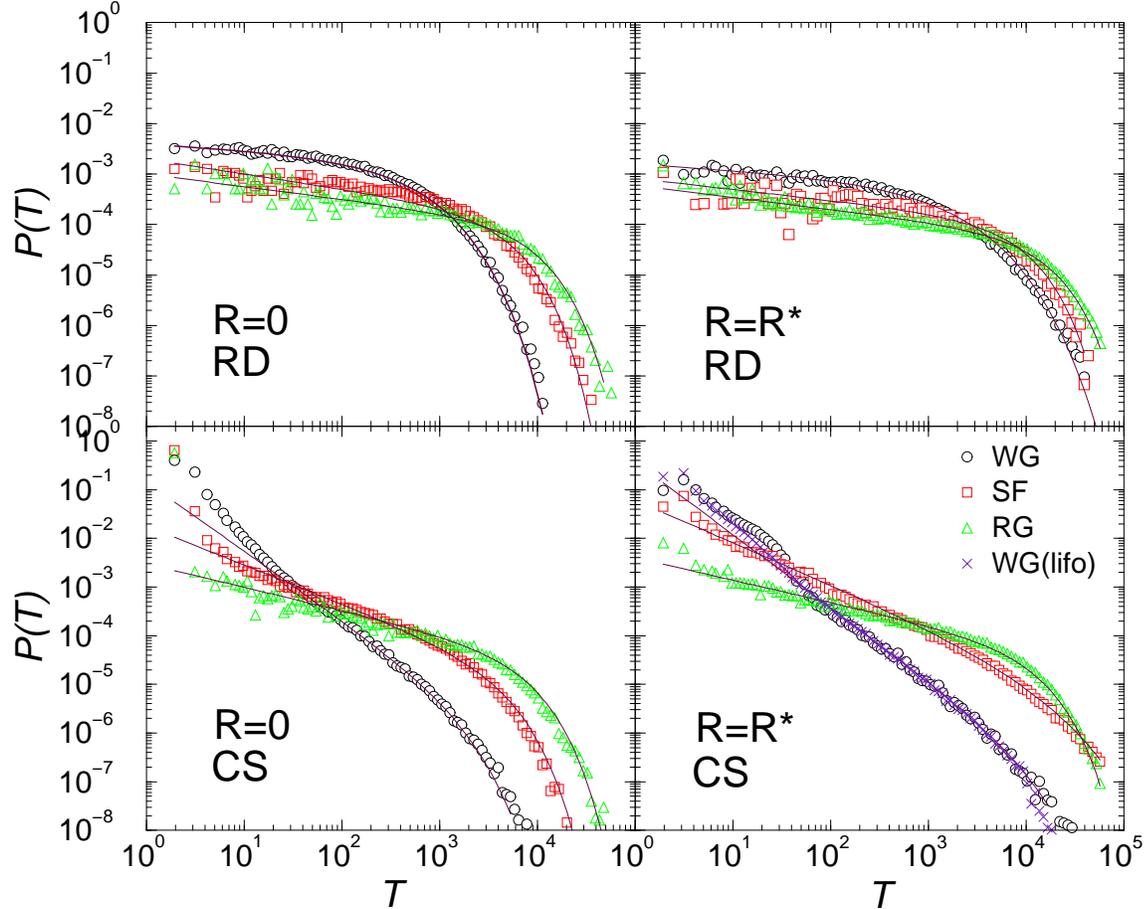}
\end{tabular}
\end{center}
\caption{
Transit times distribution $P(T)$ for
no queuing $R\to0$ (left column), and stationary 
flow at respective rate $R=R^\star$ (right column) for the  
three graph topologies. Lower panels show results  
for the CS algorithm, upper panels for the random diffusion (RD). 
The distributions are normalized by the total number of arrived marked 
packets ($2\times 10^5$ packets) and are log binned. 
Solid lines are fits to Eq. (\ref{fit_TT}).  
Corresponding parameters are given in Table \ref{Tab_fits}. 
}
\label{Fig_transittime}
\end{figure}

Consider the $R\to 0$ limit, where queuing is absent, rendering all 
waiting-times $t_w(i)=1$.
Hence the transit-time of a packet is given by the number of jumps  
before reaching its destination. 
In this limit transit-time probes
network topology and efficiency of the applied search on that topology.
In Fig. \ref{Fig_transittime} (left column) 
we show the transit-time distribution for the 
three network topologies WG, SF, and RG, and for the 
two search algorithms RD and CS. 
The results suggest that topology strongly 
matters when the advanced local search CS is applied, resulting in drastically 
shorter transit-times, especially for  the WG graph. 
Differences are also seen between the SF and RG tree graphs. 
For RD differences between the three network topologies become less 
pronounced. 

A prominent feature of the transit-time distribution is 
the power-law behavior before a cut-off. This is most apparent for 
CS. The distributions  are fitted to 
the following  expression
\begin{equation}
P(T) = A \, T^{-\tau _T}\exp{\left[-\left(\frac{T}{T_0}\right)^\sigma \right] } \quad, 
\label{fit_TT}
\end{equation}
to account for the power-law and for finite size cut-offs.  
Results are gathered in Table \ref{Tab_fits}. 
For CS the scaling exponents $\tau _T$ vary considerably 
with network topology from $\tau _T\approx 1.5$ for the WG, to 
$\tau _T\approx 0.5$ for the RG. 
For RD the cut-off time $T_0$ increases, in
agreement with lower transport efficiency. A small slope 
$\tau_T$ close to  $0.25$ was found for both tree networks and almost no
power law in the WG.
The stretching in the exponential cut-offs is absent in  most of the fits, 
however,  for CS 
on the WG fits improve with a  $\sigma =0.6-0.7$.
\begin{table}
\begin{center}
\begin{tabular}{|c|ccc|ccc|ccc|ccc|}
\hline 
\multicolumn{1}{|c}{}& 
\multicolumn{6}{c|}{random diffusion (RD)} & \multicolumn{6}{c|}{advanced {\it nnn}-search (CS)} \\
\hline 
\multicolumn{1}{|c}{}& 
\multicolumn{3}{c|}{$R=0$} & \multicolumn{3}{c|}{$R=R^*$} &
\multicolumn{3}{c|}{$R=0$} & \multicolumn{3}{c|}{$R=R^*$} \\
\hline 
Measure &RG &SF &WG &RG &SF &WG &RG &SF &WG &RG &SF &WG \\
\hline 
%$R^*$ &-    &-    &-    &0.005 &0.001 &0.005  &-    &-    &-   &0.01  &0.005 &0.2 \\
$A$ &0.001  &0.002  &0.004 &0.0006 &0.008  &0.003 &0.003   &0.018  &0.12 & 0.004 &0.06 &0.38   \\
$\tau _T$ &0.25&0.30   &0.12  &0.24   &0.22   &0.12  &0.48    &0.80   &1.42 &0.46   &0.86 &1.5    \\
$T_0$    &7000&3900   &360   &12000  &7000   &1000  &6000    &3600   &1500 &10000  &10000&10000  \\
$\sigma $& 1& 1& 0.7 & 1& 1 & 0.6 &1& 1& 1&  1& 0.6& 1\\
$q$   &1.49 &1.38 &1.51 &1.53 &1.46 &1.49 &1.42 &1.81 &-   &1.64 &1.87 &-    \\
$\beta$  &-    &-    &-    &1.99 &2.00 &1.98 &-    &-    &-   &1.99 &1.85 &1.35 \\
$\alpha$ &-    &-    &-    &0.49 &0.49 &0.45 &-    &-    &-   &0.47 &0.39 &0.21 \\
\hline 
\end{tabular}
\end{center}
\caption{Fit parameters for the fit of transit-time distributions $P(T)$, 
as defined in Eq. (\ref{fit_TT}). Also shown are the values for non-extensivity parameter $q$ defined in Eq. (\ref{sfp}) and the scaling exponents 
for the power-spectrum $\beta$ and the profile fluctuations 
$\alpha $,  of workload timeseries on different graphs.  }
\label{Tab_fits}
\end{table}

\subsection{The $R=R^{\star}$ case}
Measurable effects of queuing are expected in the stationary 
flow at  $R = R^\star $ on different networks. 
We monitor waiting-times of labeled packets in detail.
On a given graph the statistics of waiting-times is sensitive both to 
the posting rate $R$ and the search algorithm. 
In stationary flow long queues are strongly suppressed  
and waiting-times are generally short in the FIFO queuing.  
The distribution of waiting-times decays rapidly with $t_w$.
(The study  of long queues at large posting rates $R>R^{\star}$ 
is left out of this paper.)
As a consequence of  waiting-times $t_w(i)>1$ in Eq.\ (\ref{Tk})
in stationary flow we find that the 
cut-offs in the distribution of transit-times increase, suggesting a 
delay in transport. However, the slopes of the transit-time distributions 
remain roughly the same as in the $R\to 0$ limit. 
Results are shown in Fig. \ref{Fig_transittime} (right column), both for
CS and RD. Fit parameters to 
Eq. (\ref{fit_TT}) are  collected  in Table \ref{Tab_fits} and  
suggest a persistence of power-laws.
It is found that for the WG with CS the queuing
discipline only affects the transit-time distribution at short times, 
whereas the asymptotic  power-law behavior persists with a unique exponent. 
We include the result for the LIFO queuing discipline on the WG for comparison,
shown on lower right panel in Fig. \ref{Fig_transittime}.    
As before, the differences between the graph 
topologies narrow up in the case of random diffusion (see upper right panel
in Fig.\ \ref{Fig_transittime}).

The deviation from the power-law at low transit-times for CS
on the WG and the  SF graphs   (see Fig. \ref{Fig_transittime}) is a 
direct consequence of the presence of hub nodes in these graphs. 
When both the origin and destination node of a packet are in the 
vicinity of a hub, e.g., up to a three-step separation, a 
direct path between them is immediately found by  
CS. The frequency of such an event is directly proportional to
the number of  hubs in the graph and 
to the number of nodes linked to these hubs.  
The excess of the fast transport is reduced at finite posting 
rate because of queuing at the hub nodes.

\subsection{Statistical regularity  and graph structure } 
As seen in Fig.\ \ref{Fig_transittime} for CS transit-time 
distributions differ markedly on graphs with different structure. 
This means that CS makes use of the underlying graph topology to accelerate 
transport, which leads to the power-law dependences with strictly larger  
scaling exponents. 
The power-law behavior of the distributions (apart from  a cut-off) 
ensures that if a stable limit stochastic process exists, it can be 
related to the graph structure.
Here we attempt to characterize the statistical 
regularity of such a process on the  WG.

The transit-time $T_k$ for each packet is a partial sum given 
in Eq.\ (\ref{Tk}).  The set of all transit-times $\{T_k\}$ is a set of 
stochastic variables both  because of the randomness of path length $k$ 
(related to topology and search algorithm) and because of the 
random increments (queuing times) along that path.
They both depend on the simulation time, i.e., when and where 
a packet has been sent, and on the posting rate. 
We now consider the limit $R\to 0$ (no queuing),
where stochasticity arises from topology and search only. 
Since  the distribution of transit-times is  a power-law, 
in theory \cite{Prohorov_book,book_spl} all transit-times 
from the set $\{T_k\}$ can be driven from the same
probability function (stable law) $G(X)$ by rescaling the argument, 
i.e., $P(T_k<X)= G(c_kX-a_k)$. Here $c_k$ is a 
space-scaling factor, $a_k$ the average value 
and $G(X) $ is the  class of stationary distributions.
Then the distribution function of  the rescaled partial sum
converges to the same stable law $G(X)$, i.e.,
\begin{equation}
G\left(k^{-1/\mu}T_k - a_k\right) \Longrightarrow G(X) \quad.
\label{FCLT}
\end{equation}
When  $c_k=k^{-1/\mu}$ and   $0< \mu \leq 2$ it is known that the cumulative 
probability density function (pdf) which  describes the stable law  is 
given by $P(X>T) \sim T^{-\mu }$ \cite{book_spl}.
In the special case for  $\mu =1/2$ 
the stable law is described by L\'evy pdf, 
(see \cite{Prohorov_book,book_spl} for details).

The observed scaling exponent $\tau _T$ close to 3/2 for the  
$P(T)$ at  large transit-times in the  WG with CS, suggests a stable 
law with $\mu \equiv \tau_T -1 \approx 1/2$. 
One may speculate about the kind of 
limit process which is at the origin of the observed statistical
regularity. A L\'evy pdf
could be certainly understood when associated with a random walk in 
which long jumps are allowed, or in terms of the 
relation of the time of the $n$th return of a centered random 
walk to the origin \cite{Prohorov_book}, 
which could  be naturally related to cycles in a graph. 

\begin{figure}[htb]
\begin{center}
\begin{tabular}{c} 
\includegraphics[width=16.5cm]{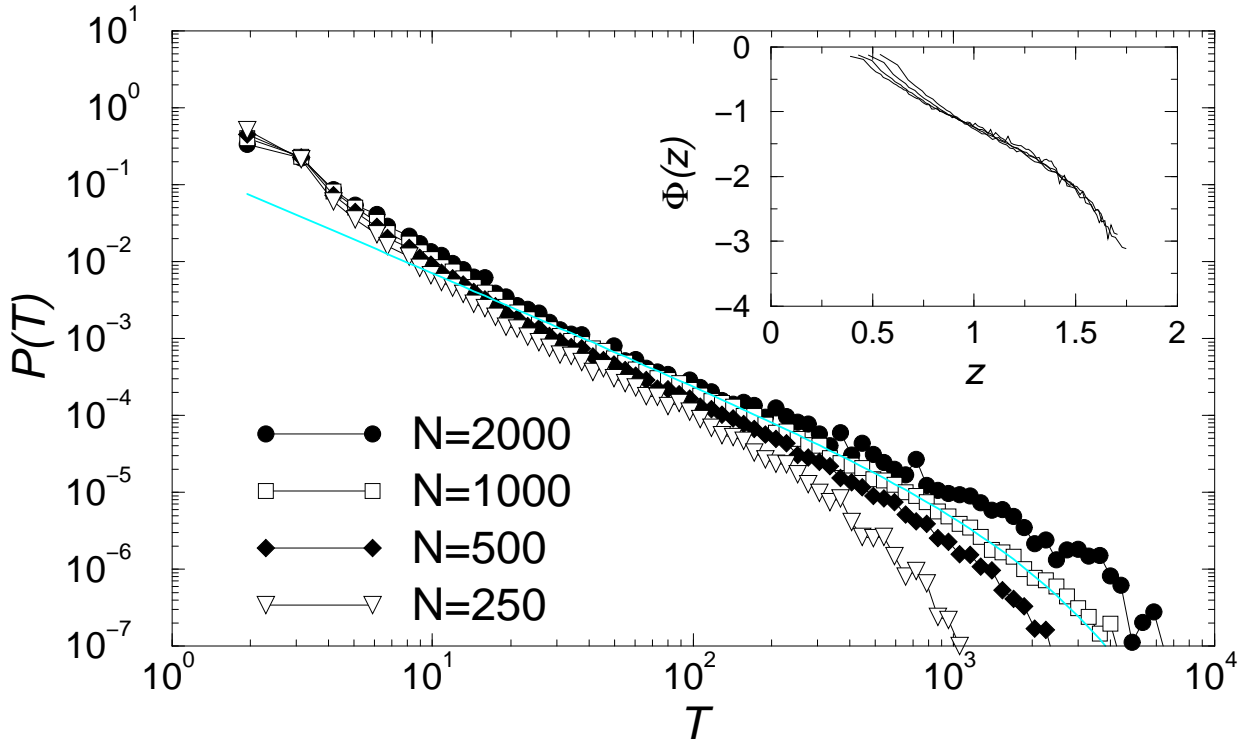}
\end{tabular}
\end{center}
\caption{Transit time distribution $P(T,N)$ vs. $T$ for  
CS on the WG in the $R\to 0$ limit for different 
network sizes $N$.  
Inset: Spectrum of dynamic exponents $\Phi (z)$
defined in Eq. (\ref{mfs_form}) with $T_0=0.1$ and $N_0=1$.
}
\label{Fig_scale}
\end{figure}

As an immediate consequence of the self-similarity
expressed in Eq. (\ref{FCLT}) one could expect the transit-time 
distribution to scale with path length $k$.  
To study this we vary the size of graph $N$. 
As stated in Section \ref{secii}, the  pairs of nodes that we  consider  
belong to the giant component of the graph. Since the giant component in the 
WG scales linearly with the graph size $N$,  we expect that the longest path 
(and thus the transit-time) will scale with  $N$. 
Simulations using different 
network sizes $N$ show that this is in fact the case 
as shown in Fig. \ref{Fig_scale}. 
An  attempted scaling fit with a unique fractal dimension (dynamic exponent) 
$z$ according to $P(T,N) = N^{-\tau_tz}{\cal{R}}(T/N^z)$ does not lead to a 
satisfactory scaling collapse and the scaling function ${\cal{R}}(T/N^z)$
can not be determined. We try a {\it multifractal} scaling fit instead: 
\begin{equation}
P(T,N) = \left(\frac{N}{N_0}\right)^{\Phi(z)} \quad; \quad
z = \frac{\log (T/T_0)}{\log (N/N_0)} \quad.
\label{mfs_form}
\end{equation}
The collapse is found when the scaling parameters $T_0$ and $N_0$ are suitably
selected. The resulting spectral function $\Phi(z)$ 
(spectrum of dynamic exponents) is shown in the inset to 
Fig. \ref{Fig_scale}. 
The range of the spectrum,  $0.5\leq z \leq 1.75$,  implies  
that the associated processes on the WG are super-diffusive 
($z < 2$) with parts of the spectrum in the 
ballistic region ($z \leq 1$). 
This can be understood as follows: As the number of nodes
in the graph increases the hub nodes become richer 
in terms of connectivity. Therefore, the probability 
of a three-link separation between selected pairs of  nodes around the hub 
increases. Transport between such nodes is 
direct and  not diffusive anymore. 
This fast transport dominates the lower part of the spectrum $\Phi(z)$ 
in large graphs. 

In the tree networks with CS and in all three graphs  with RD the scaling 
exponents (Fig.\ 3) are found in the range 
$\tau _T \lesssim 1$. These exponents  would lead to values 
$\mu \lesssim 0$, which can not be related to true self-similar 
limit  process implied by Eq.\ (\ref{FCLT}).
However, the tree network transit-time distributions 
(Fig. \ref{Fig_transittime}) also imply statistical regularity with
 non-Gaussian pdf. 
The super-diffusive nature of traffic in these cases can be discussed 
more conveniently by considering the velocity distributions 
associated with transit-times on these graphs.   

\section{Velocity distributions and $q$-statistics \label{seciv} }
We start with a short summary of the concept of non-extensive entropy,
which offers an analytical  way to solve a certain class 
of non-linear diffusion equations.   
In general, modeling non-linear anomalous diffusion processes involves
non-linearities in the associated Fokker-Planck like equations 
\begin{equation}
\frac{\partial}{\partial t} p({x},t) = 
 -\frac{\partial}{\partial x} \left(F\left[p\right(x,t)] p(x,t)\right)
+ \frac{1}{2} \frac{\partial^2}{\partial x^2}\left(D\left[p\right(x,t)] p(x,t)
\right)
\quad .
\label{fp}
\end{equation}
In the following we neglect the flow-term for simplicity 
$F[p]=0$. 
To incorporate the idea of anomalous diffusion, the functional 
dependence of the diffusion parameter in the form  $D[p]p = p^{\nu}$ 
is considered. 
With this choice Eq. (\ref{fp}) can be solved using 
the non-extensive entropy approach \cite{tsallis88,tsallis98}, 
where  classical entropy definition $S=-\int p(u) \ln p(u) du$
is modified to 
\begin{equation}
S_q=\frac{1-\int  p(u)^q du}{q-1} \quad. 
\end{equation}
By maximizing $S_q$ while keeping energy fixed 
it can be shown that the resulting $p(u)$, i.e. 
\begin{equation}
p_q(x,t)=\frac{\left[1-\beta(t)(1-q)[x-\langle x \rangle(t)]^2\right]^{\frac{1}{1-q}}}
 {Z_q(t)}
\end{equation}
also solves Eq. (\ref{fp}) without the flow-term \cite{tsallis96,bologna00}. 
Here $Z_q(t)$ is the generalized $q$-partition function.
The relation of the Tsallis entropy factor $q$, and 
$\nu$ in Eq. (\ref{fp}) is given by $q=2-\nu$ \cite{tsallis96}. 
Note, that this distribution is a power law, and only 
in the limit $q \longrightarrow 1$ the classical Gaussian 
$p_1(x,t)= \frac{
\exp^{-\beta(t) \left[x-\langle x \rangle(t)\right] ^2}}{Z_1(t)} $
is recovered. 
It can be shown that the corresponding velocity 
distribution of the associated particles is given by \cite{silva98}, 
\begin{equation}
p(v)=B_q\left[1-(1-q)\frac{v^2}{v_0^2}\right]^{\frac{1}{1-q}} \ ,
\label{sfp}
\end{equation}
which reduces to the usual Maxwell-Boltzmann result 
$p(v)=\left( \frac{m}{2\pi k T}\right)^{\frac{3}{2} }
      \exp^{-\frac{\beta mv^2}{2}}$ 
again in the limit $q \longrightarrow 1$. 
Here $v_0^{-2}\equiv \frac{\beta m}{2}$ can be seen 
as an inverse  mobility factor, meaning that large 
values of $\frac{\beta m}{2}$
are associated with small average particle velocities.   
If diffusion on networks is in fact describable as anomalous diffusion 
governed by a Fokker-Planck equation like Eq. (\ref{fp}) with the 
special choice of $D\left[p\right]$, 
the velocity distribution function should follow the general form given in 
Eq. (\ref{sfp}).

\subsection{Defining velocity on networks} 
On the networks discussed in this paper we have not 
selected a particular length scale. Moreover, the packets/particles 
which are sent along the links of the network do not 
have a particular velocity. In case a packet  is not trapped 
at a node it moves at a rate of 
{\it one link per one time step}.  
As discussed above, a sensible measure which captures the performance of a 
network with given topology and search algorithm is 
the transit-time $T$. 
Finite transit-times are governed by two effects: 
(1) The minimum distance of a given pair in terms of 
the number of links between them, and the deviation of the actual path 
from the minimum separation; 
(2) the waiting-times $t_w(i)$ at  node $i$ 
along the path taken by the packet. 

We consider the minimum distance $d$ between pairs of nodes 
picked at random on a given graph, 
which serves as a ``linear'' scale of a graph.
The velocity of a packet transported
between a given pair can be defined as the ratio $v = d/T$. 
The distribution of distance over random 
pairs carries information about graph topology.  
\begin{figure}[htb]
\begin{center}
\includegraphics[width=16.5cm]{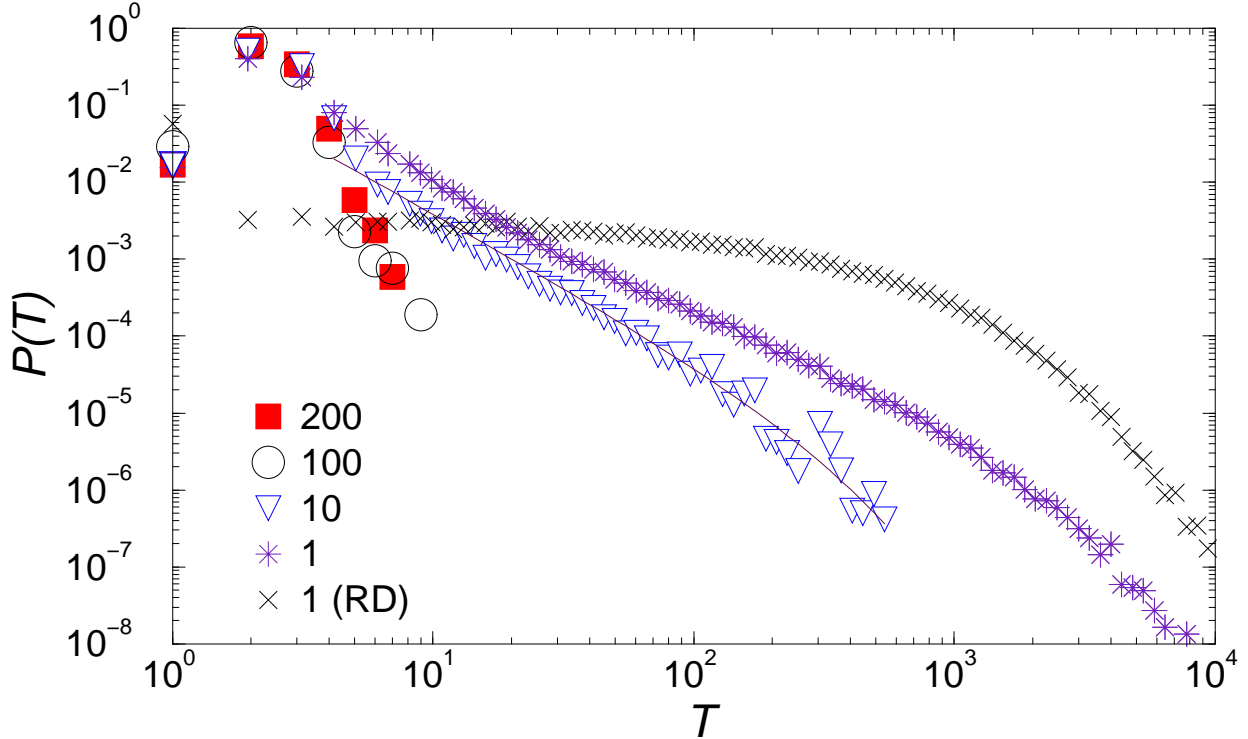} \\
\end{center}
\caption{
Probability distribution of transit-times $T$  
corresponding to shortest separation of pairs in the WG  (full squares).  
Different curves are for the shortest transit-times  of 
200 (squares), 100 (circles), 10 (triangles) and 1 (*) identically sent packets 
between the same pair of nodes in the $R\to 0$ limit for CS rule.
The case for a single copy sent with  RD search is also shown as (X). 
}
\label{Fig_minpath}
\end{figure} 
Note, that in the $R\to 0$ limit
the distribution of minimum path lengths corresponds one-to-one 
to transit-times of packets if a {\it global} optimal navigation is applied. 
Here we determine the distribution of shortest separations statistically, 
i.e., as the stable statistics with a large number of trials.
For the WG this is a narrow distribution sharply peaked at 
an average distance $\bar{d} \geq 1$ (Fig. \ref{Fig_minpath}). 
The minimum path is limited by the diameter $\bar{D}$ of the graph
$d\leq \bar{D} \propto \log (N)$, which is very small in the structured graphs. 
Therefore, as a good measure of velocity of packets on the 
graph we may take $v \sim 1/T$, where $T$ is the actual transit-time of a 
packet.

\begin{figure}[htb]
\begin{center}
\begin{tabular}{c} 
\includegraphics[width=14.5cm]{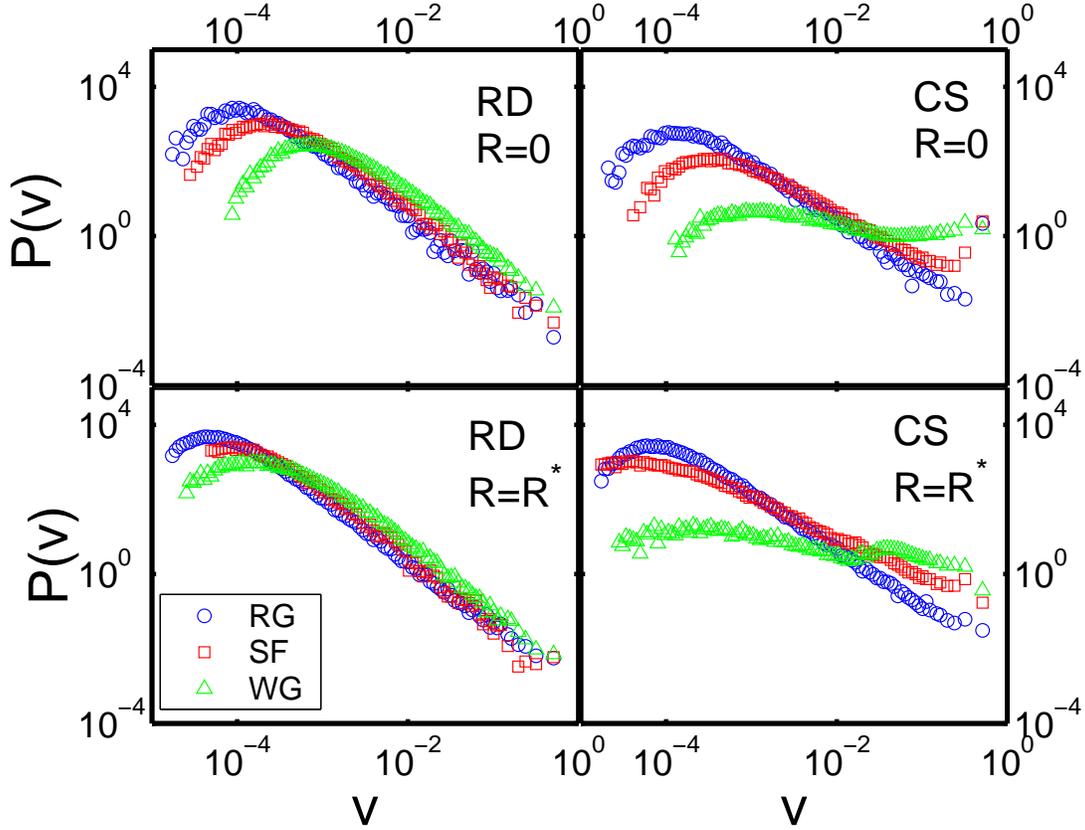} \\
\end{tabular}
\end{center}
\caption{Velocity distributions $p(v)$ for the three  
networks, for $R\to0$ (top row), and $R=R^*$ (bottom).  
RD is shown in the left column, CS in the right.}
\label{Fig_v_distrib}
\end{figure}

The probability distribution of the velocity of traffic at each graph can be
either computed directly in the simulations or derived from the 
transit-time distributions. 
The situation in  both approaches is qualitatively the same. We 
get somewhat smoother curves for the derived distributions.
By noting $p(v)dv =P(T)dT$ we straight forwardly transform 
\begin{equation}
p(v) = P(T) \left| \frac{dv}{dT} \right|^{-1}  =P(1/v)v^{-2} \quad .
\label{v_dist}
\end{equation}
Since the velocity on a graph is by our definition  
a ratio of stochastic variables,
one could expect the probability density function 
of the form (apart from the
finite size cut-offs)
\begin{equation}
p(v) =  \frac{\sigma}{\pi (\sigma ^2 + v^2)^y} \quad
\label{velocity_pdf}
\end{equation}
where  $y\neq 1$ measures deviation from the standard Cauchy pdf.
The case $y=1$ corresponds to the pdf of the ratio of two 
IID variables.
In  Fig. \ref{Fig_v_distrib} we show the velocity distributions 
obtained from  the transit-times of Fig. \ref{Fig_transittime} 
for different graph types. For RD 
the slope of the velocity distributions  on all three graph structures is 
$\tau_v \lesssim -2$. The deviations from $y=1$ are clearly 
linked to the network topology when CS is used. 
We suggest to use these deviations as a measure of correlation 
for a given search algorithm with topology.
  
As a generic feature, $p(v)$ has power law tails
which extend over several decades. For RD
these tails extend to the very end for all networks, 
while for CS the WG shows a peculiar behavior: 
For high velocities the probability density 
moves up again. This is in agreement with increased  probability 
of  fast transport in the close vicinity of the hubs,
as discussed in Section \ref{seciii}.  

The distribution $p(v)$ in  Eq. (\ref{v_dist}) for
various networks for different posting rates 
and search algorithms can be  compared to 
the form of Eq. (\ref{sfp}). From fits to this
curve the value of $q$ can be estimated. 
Technically, this was done by linear fits to the 
so-called $q$-logarithm,  $q\_ \log p(v) =
\left[1 - p(v)^{1-q} \right]/(q-1)$. 
For the appropriate value of $q$ this 
expression becomes a linear function of $v$. 
$q$ was obtained by finding the minimum of the 
least square errors of linear fits to  
$q\_ \log p(v)$ as a function of $q$. 
Results are gathered in Table \ref{Tab_fits}. 

Values of $q > 1$ are characteristic for the super-diffusive nature  
of the transport on these graphs \cite{tsallis96}. 
The velocity distribution in the WG in combination with  CS 
has a much lower power exponent which would imply un-physically 
large $q\gg 2$. 
This is in agreement with the multifractality of transit-times on the WG, 
discussed in the previous section. A single value of $q$ probably can not 
properly describe the
character of the process which has the spectrum of the dynamic exponents
$\Phi(z)$ shown in  Fig. \ref{Fig_scale}. 

Note, that the velocity pdf in Eq. (\ref{velocity_pdf})
is equivalent to the general  form in Eq. (\ref{sfp}), when  
parameters are identified as 
$y \equiv 1/(q-1)$, $\beta m/2 \equiv 1/\sigma ^2 (q-1)$, and
$B_q\equiv \sigma ^{(q-3)/(q-1)}/\pi.$
The non-extensivity parameter $q$ in the  anomalous diffusion
on networks can be directly related to the network structure and 
correlation between search and topology, indicated by $y \neq 1$.

\section{Workload timeseries \label{secv} }
Another source of information is the workload timeseries $n(t)$ 
where we look for non-trivial correlations (Fig. \ref{figrate}).  
Self-similarity of the process and related persistence in a
timeseries $X_t$ 
can be characterized by the temporal correlation function 
of its increment process $I_t=X_{t+1}-X_t$, 
\begin{equation}
 C(\tau) =  \langle I_t I_{t+\tau} \rangle =  
\frac{1}{N-\tau} \sum_{t=1}^{N-\tau}  I_t I_{t+\tau}
 \quad. 
\label{correlator}
\end{equation}
For no persistence, i.e., a temporally uncorrelated $I_t$ process,   
$C(\tau)=0$ for $\tau\leq 0$. For persistence up to 
$\tau_{\rm pers}$ timesteps $C(\tau)$ is positive below 
$\tau_{\rm pers}$,  and vanishes above. 
Reliable direct computation of $C(\tau)$ is known to be problematic, 
due to noise, possible non-stationarities or trends and the 
finiteness of the  timeseries \cite{thurner97}. 
We therefore adopt a standard way of proceeding 
\cite{bunde98}, and start from the 
original timeseries $X_t$, 
interpreting it as the position of a  random walker. 
$X_t$ is sometimes called a profile.
It is known that for a power-law decay in the correlator  
\begin{equation}
C(\tau) \sim \tau^{-\gamma}, \quad 0 < \gamma < 1 \quad, 
\label{correlator_scale}
\end{equation}
the statistics (standard deviations) of the profile-fluctuations, 
defined by 
\begin{equation}
F(\tau) = \langle 
( X_{t+\tau} - X_{t} )^2\rangle_{t} ^{\frac{1}{2}} \quad ,
\end{equation}
also scale according to 
\begin{equation}
F(\tau) \sim \tau^{\alpha}, \quad \alpha= 1-\gamma/2  \quad. 
\label{correlator_F}
\end{equation}
There are several possibilities to define $F(\tau)$, which 
lead to the same (or one-to-one related) scaling behavior. 
This can be used to eliminate the influence of linear 
trends in the data. Frequently used definitions are 
''detrended'' fluctuation analysis,  
or statistics based on wavelet analysis \cite{arneodo95,thurner98}. 
These methods are superior to the direct analysis of 
the power-spectra since non-stationarities 
in the data are known to influence the quality of scaling 
exponents. However, the power-spectrum density, 
where squared Fourier transforms $S(f)$ of the data  are considered, 
has the advantage of a  straight forward interpretation: 
It directly gives the energy distribution of the different 
frequencies present in the system. For this reason we include the 
power-spectral density in this analysis. Starting  
from the workload profile $X_t$, one can expect for scaling processes 
\begin{equation}
 S(f) = 
\left| \sum_{n=0}^{N-1} X_t \,\, e^{i \frac{2 \pi t(f-1)}{N} }  
\right| ^2  
\propto f ^{-\beta} \quad.
\label{psd}
\end{equation}
For existing power-law decays in the correlations $C(\tau)$, 
the relation $\beta = 3 - \gamma=2\alpha+1$ holds. This relation is often 
only poorly recovered in realistic data, due to artifacts originating from 
noise, trends and finite data size.  
The scaling  $\alpha$ and $\beta$ serve to characterize the 
process in terms of temporal correlations: 
While for classical Brownian motion ($\alpha=0.5$, $\beta=2$) there are no 
correlations, for $\alpha>0.5$ or $\beta>2$ the process is called persistent, 
i.e., if the process was moving upward (downward) at 
time $t$ it will tend to 
continue to move upward (downward) at future times $t'>t$ as well. 
This means that increasing (decreasing) trends in the past 
imply -- on average -- increasing (decreasing) trends in the future.  
If  $\alpha<0.5$  the corresponding process is called anti-correlated 
or anti-persistent, meaning that increasing (decreasing) trends in 
the past imply -- on average -- decreasing (increasing) trends. 
\begin{figure}[htb]
\begin{center}
\begin{tabular}{cc} 
\includegraphics[width=7.4cm]{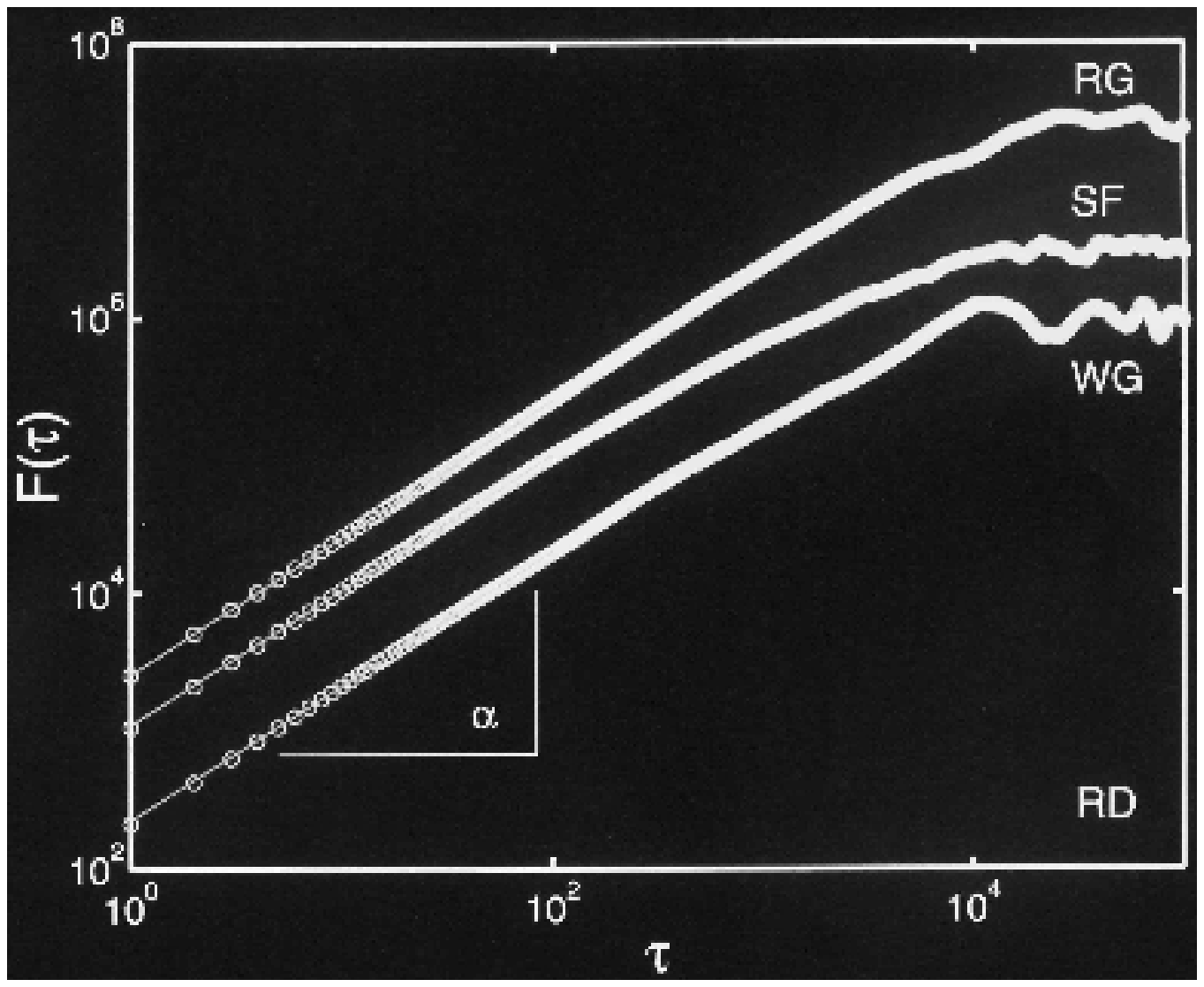} &
\includegraphics[width=7.4cm]{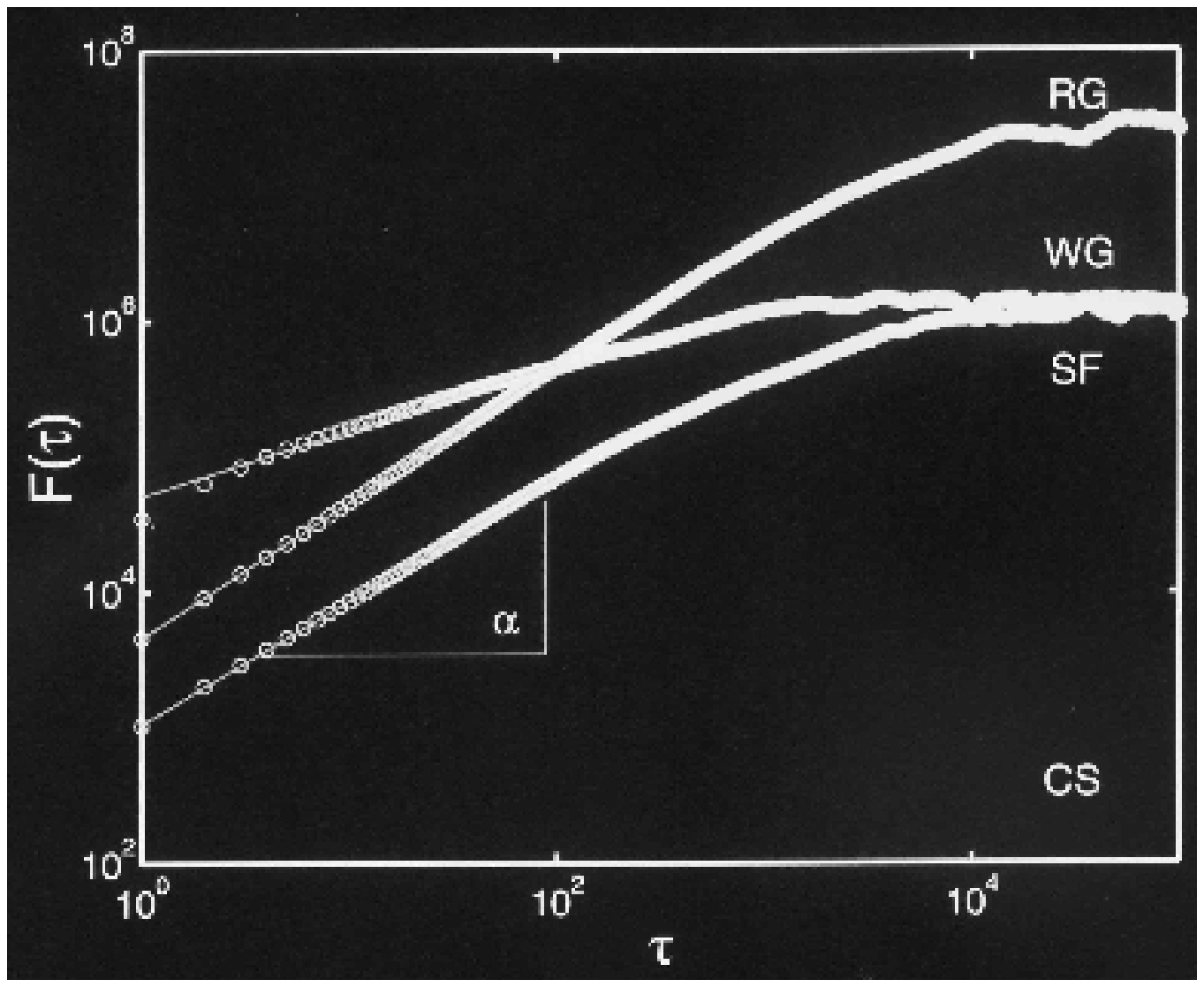}  \\
\includegraphics[width=7.5cm]{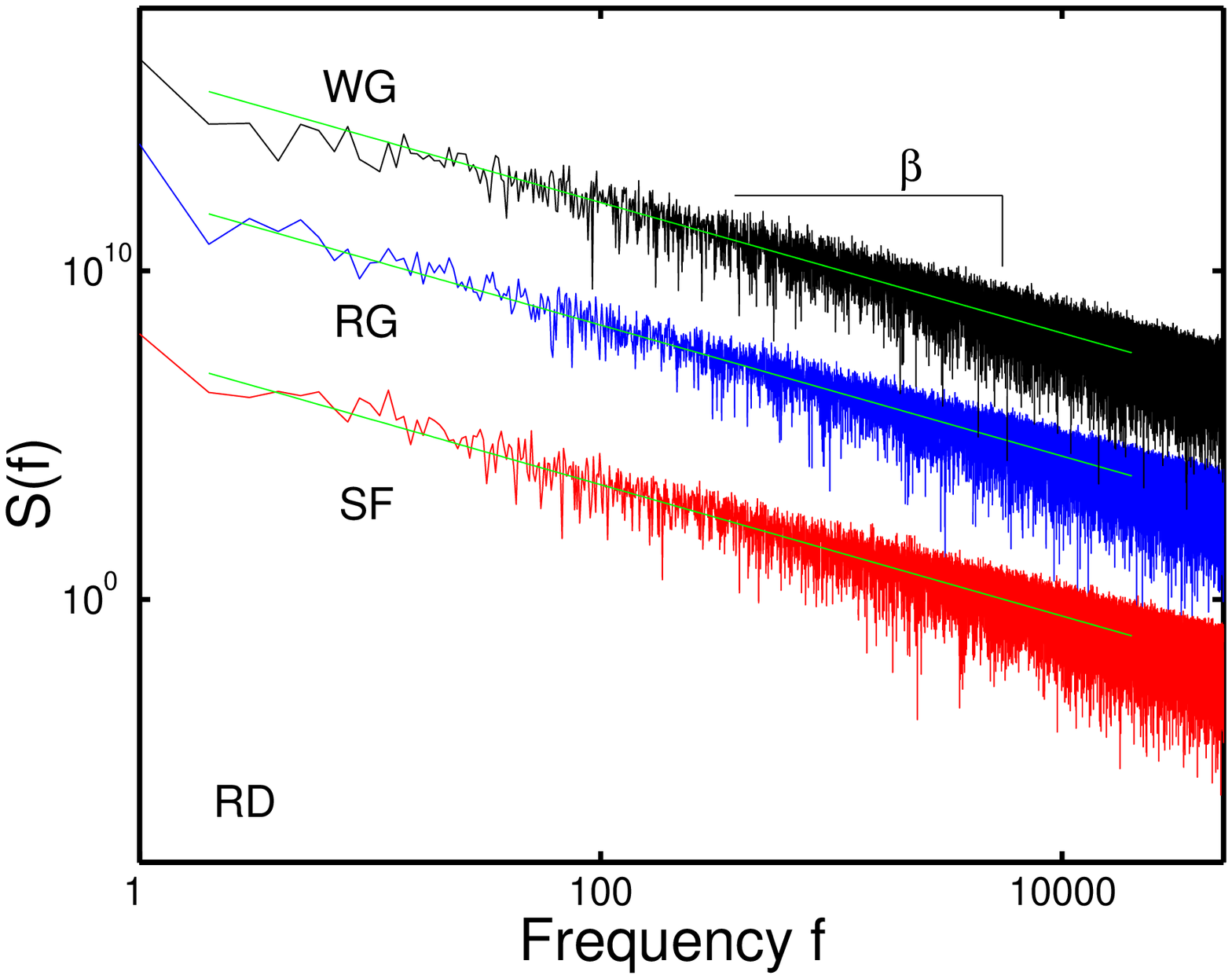} &
\includegraphics[width=7.5cm]{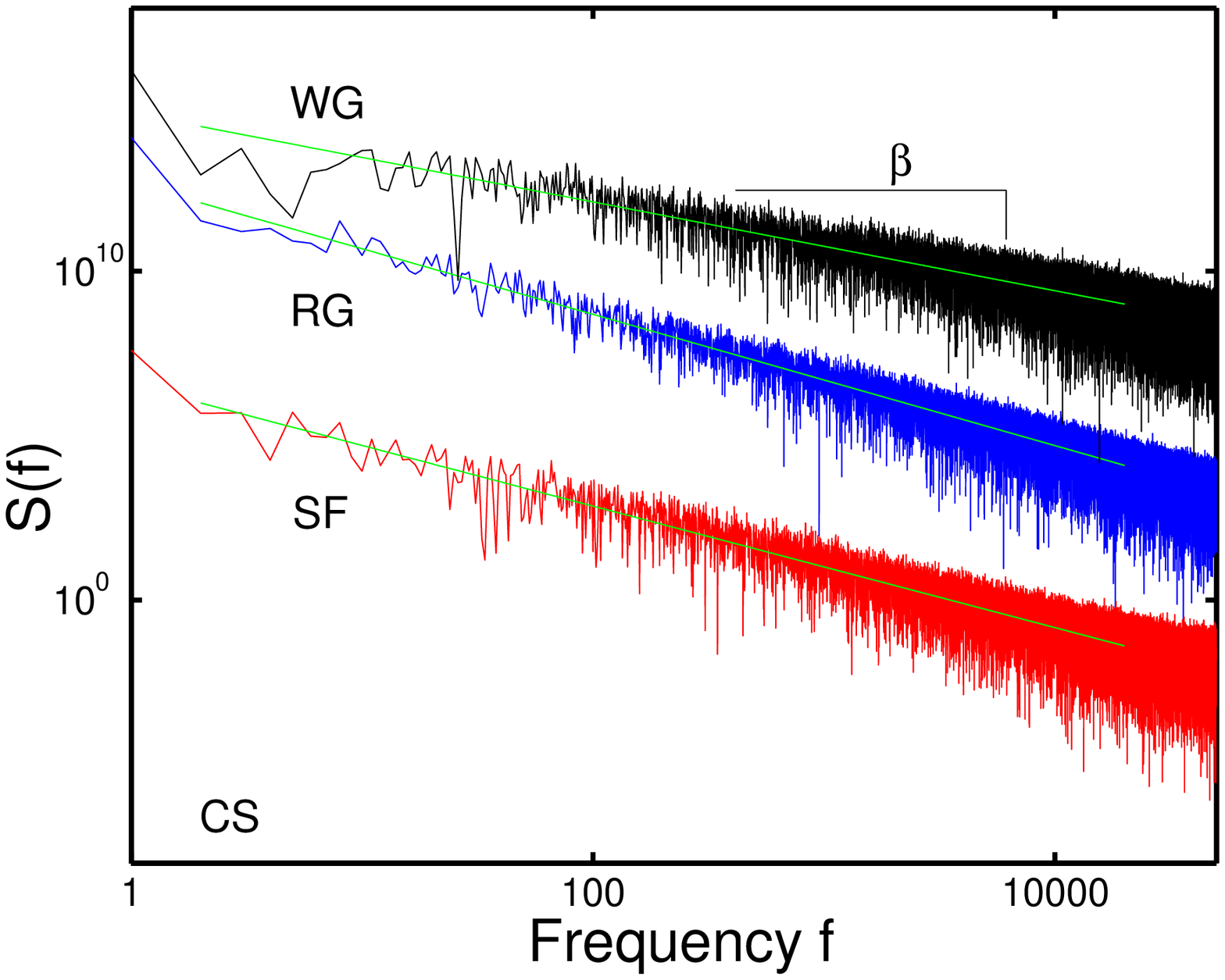} \\
\end{tabular}
\end{center}
\caption{Correlation functions $F(\tau)$ defined in Eq. (\ref{correlator_F}), 
and power-spectra of the 
$1/f^\beta$ type from Eq. (\ref{psd}), 
in stationary flow at $R=R^\star$ 
for different network topologies and diffusion rules: random diffusion (RD) and 
{\it nnn}-search (CS).
Corresponding exponents are gathered in Table \ref{Tab_fits}. 
Power spectra have been shifted horizontally for different graphs.
}
\label{Fig_spectra}
\end{figure} 

In Fig. \ref{Fig_spectra} correlators $F(\tau)$ and power-spectra 
$S(f)$ are shown for stationary flow at $R=R^\star$ for the three 
graph topologies and diffusion rules. 
The corresponding scaling exponents have 
been obtained by least square fits to the data. 
Results are found in Table \ref{Tab_fits}. 
For RD both $\alpha$ and $\beta$ suggest practically 
uncorrelated workload processes, regardless of the network type. 
The situation changes for CS 
where signs for anti-persistence are found 
in the  SF ($\alpha=0.40$,  $\beta=1.85$) and especially in the 
WG ($\alpha=0.21$,  $\beta=1.35$). 

Observed anti-persistence ($\alpha < 1/2$ ) may be related to some kind
of regulatory effects due to powerful hubs in the structured networks. 
As discussed above, the advanced 
search often directs packets via hub nodes,  
and even for low global 
workload, there can emerge long queues at large hub nodes. 
While waiting in a queue, packets increase the network's workload, but they 
do not contribute to randomness of the process. 
Moreover, when a waiting packet is on turn to be processed, it gets quickly 
to the target (which
is often found in the neighborhood of that hub) and 
disappears from the traffic; In other words, 
the system corrects itself by quickly decreasing 
the workload.  Connected with this scenario is the 
fact that networks with more powerful hubs, such as the WG, 
are more efficient than  networks with only one hub 
(e.g. SF) or no hubs at all (e.g. RG).  

\section{Discussion \label{secvi} } 
In this work by use of large-scale numerical simulations and theoretical 
concepts we demonstrated how packet transport between nodes on  
different structured 
networks can be quantified as an anomalous diffusion process. 
In our  model we control both graph topology and
microscopic diffusion rules. Our main aim was to explore how topology 
of graphs affects properties of the diffusion process. 
For this purpose we ensured 
a stationary un-jammed flow on each graph topology by an  
externally fixed posting rate $R$. 

We based our work primarily on the analysis of transit-times of packets.  
We find power-laws in the transit-time distributions which 
imply  non-negligible probabilities for very long delays. 
We find a marked difference in power-law exponents depending 
on graph topology and search algorithm. 
These power-laws are rather independent of the
posting rate, suggesting a statistical regularity of the diffusion process
that can be related to graph topology when below the jamming transition. 
For the range of posting rates studied, we find  that the 
particular queuing discipline is of minor importance. 
The type of queuing is known to play a more important role
for larger posting rates where waiting-times of packets increase when 
approaching the jamming transition.
Our main conclusions which characterize the diffusion process
on structured networks can be summarized in the following points:

\subsection{Stationarity} 
We have demonstrated that the traffic of 
packets on structured networks can be made stationary without 
imposing any global
constraint. We find the stationary jammed flow for a range of values for  
the posting rate which balances the output rate (number of packets delivered 
per time step). The output rate strictly depends on the network's 
topology and the search algorithm to navigate packets. 
Generally, for local search
algorithms more structured networks, e.g., networks with more powerful hubs
and closed cycles have larger output rates. Theoretically, the stationarity of 
the traffic guarantees that we are dealing with self-similar stochastic 
processes, which have major advantages for theoretical analysis 
(scaling). 

\subsection{Super-Diffusivity} 
We found  that 
the traffic of information packets is super-diffusive 
on any non-trivial graph
topology (emergent topologies in evolving  networks). In formal terms, 
this means that distances between nodes in a large graph can be traversed in
times that scale sub-quadratically with the graph size 
(the dynamic exponent $T\sim N^z$ is $z < 2$).
This conclusion remains  true even if packets are left 
to diffuse randomly.
This is in marked contrast to random diffusion on 
space-filling structures, where 
random diffusion is represented by Brownian motion.  
Another finding is that in the free-flow regime, 
differences between graph topologies
are marginal when packets diffuse randomly. This underlines the  
necessity for improving and adapting local diffusion rules in order to 
optimally use the topology
of an underlying graph. To this end we studied transport with the improved 
local search algorithm which looks for the destination node in 
next-near-neighbor surrounding of the node.

\subsection{Efficiency} 
When using the improved local search, transport remains 
super-diffusive but marked quantitative differences due to graph 
topologies emerge.
The improved local search in combination with some network 
topologies performs strikingly  more  
efficient than for others.  This is especially true 
for highly structured networks. The efficiency 
of the network with improved local search is directly
related to regulatory effects 
played by large hub nodes, in particular by the 
number of hubs and the number of closed cycles. 
Although the distributions of transit-times of packets in all structured 
networks are markedly different (generally shorter 
times in highly structured networks), they still exhibit power-tails. 

The optimum strategy for diffusion on a network 
would be the knowledge of the shortest paths
between the initial point and its destination. This 
knowledge would however require costly global navigation, 
which is often not available in real 
networks. Traffic along geodesics 
may also be inconvenient in the case of 
queuing, given that in the scale-free graphs a majority of shortest 
paths pass through hub nodes, while other nodes carry much less traffic. 
To understand this better we mention that the probability distribution 
of the number of paths through a node is a fast decaying 
function $P(u)\sim u^{-2.0}$ in SF and $P(u)\sim u^{-2.27}$ in WG 
\cite{BT_Granada}.
An advanced local search such as CS is certainly better in this 
respect, since it disperses the 
activity over more nodes in the network. 
This is achieved through the random part
of the search, i.e., when the destination node appears 
not to be in the locally 
searched nnn area. This feature of the local 
search implies, however, that some packets may get stuck 
in remote areas of the network, leading 
to the long tail in the transit-time distribution.

To improve  search algorithms on structured graphs 
without extending their local character one can think of a 
statistical improvement which might be realized in many 
natural systems. To this end one  
sends multiple copies of the same packet to the same destination.  
Since all packets move independently and randomly  
there would be differences between path lengths 
selected by the different copies. We monitor the shortest transit-time 
and ignore later arrivals. 
The first-arrival statistics of  
multiple  copies of the packet is summarized in Fig. \ref{Fig_minpath}.
It shows that a reasonable improvement is achieved already by sending ten
copies of the packet. Statistically, the 
fastest arriving packets use the minimum path 
up to four-link separation and longest transit-times are reduced by  an order 
of magnitude compared to the one-copy case.  
The asymptotic tail has the slope close to 1.9.
Similar effects -- reduced transit-times by first arrivals -- 
are found in the other two graph topologies, with the slope 
of the curve practically unchanged,  compared to the  
one-packet distribution.
The efficiency of information transport on 
biological networks -- e.g., signaling
molecules sent by  a node to a particular 
destination where they can bind chemically -- probably 
utilizes the above statistical methods for  improvement. 
Many identical molecules are being sent from an area 
of production (e.g. the nucleus) to a target position 
via a network (cytosceleton).The first arriving molecule binds and 
chemically acts on the target node. In technological 
networks the statistical improvement method
may  be used in low density traffic. 
For instance, in the stationary flow regime 
on the Web graph with CS at rate $R\leq 0.04$ one can copy each packet by a 
factor of ten  without yet causing jamming on the network. 
Other topologies are more sensitive to 
jamming and  the number of copies
needs to be adjusted to the basic driving rate $R$.
 
Maybe our most interesting result is that 
in several measures (transit-times, $q$ and correlation 
exponents) we found that neither 
graph topology nor the search algorithm alone are 
sufficient to explain for a drastic increase of efficiency 
of network traffic. We rather find that the combination 
of both -- in form of a good matching -- 
are the source of this improvement. 
It is most intriguing that for ordinary diffusion 
very little dependence of network structure was found, 
meaning for a unguided diffusion 
no advantage from  more structure in a network is gained.  
In other words either the 
structure of the network should be adjusted to 
the processes that the network supports, or the network 
has to be formed to use these processes most effectively.
With this finding it becomes possible to  
speculate that evolution has adopted certain 
network structures in nature to cheap and robust search mechanisms, 
and/or has adapted search strategies to existing 
networks.  

%%%%%%%%%%%%%%%%%%%%%%%%%%%%%%%%%%%%%%%%%%%%%%%%%%%%%%%%%%%%%%%%%

\end{document}